\documentclass[aps,prd,twocolumn,groupedaddress]{revtex4}
\usepackage{amsmath}

\begin{document}

\title{Semiclassical Quantum Computation Solutions to \\ the Count to Infinity Problem: \\ A Brief Discussion }

\author{Bur\c{c} G\"{o}kden}
\email[]{e119744@metu.edu.tr}

\affiliation{Middle East Technical University
\\06531, Ankara, Turkey}

\date{May 26, 2003}

\begin{abstract}
In this paper we briefly define distance vector routing
algorithms, their advantages and possible drawbacks. On these
possible drawbacks, currently widely used methods split horizon
and poisoned reverse are defined and compared. The count to
infinity problem is specified and it is classified to be a halting
problem and a proposition stating that entangled states used in
quantum computation can be used to handle this problem is
examined. Several solutions to this problem by using entangled
states are proposed and a very brief introduction to entangled
states is presented.
\end{abstract}


\maketitle

\section{Introduction}

Distance vector routing is a popular dynamic routing algorithm
which is used in many applications due to its simplicity and ease
of implementation. Although it is not a technically superior
algorithm, its availability even before it was standardized made
it the most common algorithm used by now. Despite such advantages
of distance vector routing algorithm which was firstly proposed by
Ford and Fulkerson \cite{bellman, fulkerson}, the algorithm has a
very critical problem embedded in it, which arises when one of the
nodes in the network goes down(or is isolated from the network).
Since distance vector routing simply depends on routing table
exchange of a node with its neighbors, the other nodes neighboring
the node that has just gone down still think that their other
neighbors have a better path leading to that isolated node, which
in turn starts an endless exchange of data between the nodes, and
it is well known as count to infinity problem.

A good standard in distance routing protocols is called
RIP(Routing Information Protocol) \cite{rfc1058} and it solves
count to infinity problem by adding more check actions and
limitations to the system, these methods are called split horizon
and poisoned reverse. Split horizon together with poisoned reverse
solves loops in the network up to and including two gateways and
if more than two gateways are in a loop the problem is not
eliminated. Poisoned reverse loads a special meaning to the
infinite distance metric (16 for RIP) and updates other nodes'
routing tables accordingly to avoid looping so that the the
neighboring nodes do get infinite metric entry into its
corresponding table to immediately prevent a loop. But this
poisoned reverse has a serious problem that it limits the
bandwidth of the system since the packets that prevent the loop
get bigger and bigger as the network neighborhood enlarges. For
this reason, RIP is suggested to be implemented in networks no
more than 15 hops(i.e., 15 gateways connecting asynchronous
networks to each other is an example). Therefore, the major ways
of preventing count to infinity problem lead to more complications
and changes in the protocol, which adds overheads in using time
and space sources (i.e., more delays and less bandwidth with no
improvement in performance but some stabilization). Moreover,
these preventing algorithms are only applicable when the network
size is small, which is due to the the time considerations in loop
detection. To propose a solution to the count to infinity problem
with minimal loss in time and space, we will approach the system
as a cause for the count to infinity problem and show that
classical computers are not capable of solving this problem which
in turn will need entanglement as a solution for the problem. Next
section will show that count to infinity problem is a very hard
problem (in view of algorithmic complexity considerations) to be
completely solved by classical computers if the case is imminent
in a certain algorithm.

\section{Reduction of counting to infinity problem to the halting
problem}

The halting problem is one of the oldest unsolvable problems of
computation theory. It stems from Hilbert's Entscheidungs
problem(decision problem) which asks whether there is an algorithm
for solution of any problem given. The halting problem is Turing's
answer to this question and it is a well known issue that,
unfortunately, all problems do not have feasible algorithms for
their solution. The halting problem can be proved by the following
discussion\cite{chuang}. Define a Turing machine $M_{x}$ which
halts upon input of a specific number $x$ so that each such Turing
machine M is related with a Turing number $x$. And define the
halting function $h(x)$ as
\begin{equation*}
 h(x) = \left\{\begin{aligned}
                  &0, ~~\text{if machine with Turing number $x$} & \\
                  &   \text{does not halt upon input of $x$} &
                  \\[2mm]
                  &1, ~~\text{if machine with Turing number $x$} & \\
                  &   \text{halts upon input of $x$}           &
                \end{aligned}
                \right.
\end{equation*}
Therefore the halting problem can be stated as follows: Does the
machine with number $x$ halt upon input of the same number $x$, or
equivalently is there an algorithm to evaluate $h(x)$? Now suppose
that we have such an algorithm denoted by $HALT(x)$ that evaluates
the function $h(x)$.Then define the function $TURING(x)$ with the
pseudocode \\[2mm]
$
\begin{aligned}
TURING(x): \\[2mm]
     &y = HALT(x)&\\
     &\text{if $y = 0$ then halt}&\\
     &\text{else loop forever}&\\
     &\text{end if}&
\end{aligned}
$ 

Since we assumed $HALT()$ is a valid program $TURING()$ is also a
valid program and for the halting function, $h(t)=1$ if and only
if $TURING()$ halts on input of $t$(Note that if $h(t)=1$, Turing
machine ends operation giving a result and the algorithm in this
way automatically ends). But from the program, $TURING()$ halts on
input of $t$ if and only if $h(t)=0$. Thus $h(t)=1$ if and only if
$h(t)=0$, which is a contradiction and there is no algorithm
$HALT(x)$ for evaluating $h(x)$.

Count to infinity problem can be also defined in the class of
problems that have the characteristics of halting problem and to
show this it is enough to show a function derived from the count
to infinity problem, which models the function $h(x)$ above. Our
$h(x)$ for the count to infinity problem can be given as
\begin{equation*}
h(x) = \left\{ \begin{aligned}
                 &0, ~~\text{if there is not a well defined} & \\
                 &\text{route to a just isolated node with}& \\
                 &\text{acceptable amount of metric used}& \\[2mm]
                 &1, ~~\text{if there is a well defined route} & \\
                 & \text{to a just isolated node with some optimum} & \\
                 & \text{finite amount of metric used}
                 \end{aligned}
                 \right.
\end{equation*}

Here the problem only covers the nodes that are on the path that
is not available just after the destination node is isolated. If
there were an algorithm to determine a finite cost path we would
then have $h(x)=1$ after some acceptable amount of time and the
system will halt with a specific route ending the program while
$TURING()$ procedure is in a loop. Vice versa, if there were an
algorithm that can evaluate $h(x)=0$, $TURING()$ procedure would
stop and a decision would be definitely made. But for the count to
infinity problem, both of those do not concur each other's
decision as in the halting problem and in fact this problem is an
open ended issue in computer science that implies no current
solution with classical computers. When one attempts to solve this
problem based on a classical computer architecture, he must either
completely change the algorithm or take some precautions not to
let it happen, which in turn adds significant overheads and
limitations to the system. Split horizon with poisoned reverse is
such a precaution that still has problems but works well in small
networks to some extent. For these reasons we will propose a new
feature coming from quantum theory, the entangled states to solve
this problem for larger networks, which change the hardware by
using a phenomenon that has no classical counterpart in physics.
The main aim of this change in hardware is to design a feasible
system that extends to distant networks with minimum overhead, to
significantly reduce the time and space complexity of the system
which is practically applicable by the users of the network.

\section{Quantum Computation Basics and Entanglement}

Quantum computation is a theory based on quantum theory and
today's theory of computation. Below are some mathematical and
physical tools that are used in this study for defining entangled
states:

1)A qubit (equivalent of a bit in today's computers) is defined as
  \begin{equation}
  |\Psi>=a|0>+b|1> \label{eq1}
  \end{equation}
  where $|0>$ and$|1>$ are orthonormal bases for this system and
  $a$ and $b$ are complex numbers satisfying the equation
  $|a|^2+|b|^2=1$ where $|a|^2$ and $|b|^2$ are the probabilities
  for the $|\Psi>$ for evaluating to 0 or 1 upon measurement of
  that qubit. As it is seen from eq. \ref{eq1} a qubit is the
  superposition of both being 0 or 1 as a value before it is
  measured by a measurement device.

  2)Any state $|\Psi>$ has a hermitian conjugate $<\Psi|$ whose
  inner product is given as $<\Psi|\Psi> = 1$ when normalized.
  Note that $<0|0>=<1|1>=1$ and $<0|1>=<1|0>=0$ and
  \begin{equation*}
  |\Psi>=a|0>+b|1> \Rightarrow <\Psi|=a^*<0|+b^*<1|
  \end{equation*}
  and
  \begin{equation*}
  <\Psi|\Psi>= |a|^2+|b|^2 = 1
  \end{equation*}

3)Define a projection operator $P$ for the bases $|0>$ and $|1>$
which has the properties
\begin{eqnarray}
P_{0} = |0><0| \\
P_{1} = |1><1| \\
P_0 |0> = |0> \\
P_0 |1> = 0 \\
P_1 |0> = 0 \\
P_1 |1> = |1> \\
P_0 + P_1 = I = \begin{pmatrix}
                     1 & 0 \\
                     0 & 1
                \end{pmatrix}\\
|0> = \begin{pmatrix}
           1 \\
           0
      \end{pmatrix} \\
<0| = \begin{pmatrix}
             1 & 0
      \end{pmatrix} \\
|1> = \begin{pmatrix}
              0 \\
              1
      \end{pmatrix} \\
<1| = \begin{pmatrix}
              0 & 1
      \end{pmatrix}
\end{eqnarray}
Note that projection operators have two eigenvalues which are
either 0 or 1.

4) For any operator $A$ the expected value of that operation on a
state is given as $<\Psi|A|\Psi> = a \in \mathbf{R}$ where a is
the expected value of a measurement by a device.

5) Tensor product $\otimes$ is used to define an ensemble of two
or more states (or qubits) with the following properties.
\begin{itemize}
  \item {For an ensemble of two states
  \begin{eqnarray*}
  |\Psi>\otimes|\varphi> = &&a|0>\otimes|0> + b|0>\otimes|1>\\ &&+ c|1>\otimes|0> + d|1>\otimes|1>
        \end{eqnarray*}
  or shortly
        \begin{equation*}
        |\Psi\varphi> = a|00> + b|01> + c|10> + d|11>
        \end{equation*}
        where $|a|^2+|b|^2+|c|^2+|d|^2 = 1$}
  \item{ For an operator $A$ that is to be applied on $|\Psi>$ and operator $B$ to be applied on $|\varphi>$ we
  have
  \begin{eqnarray*}
  (A\otimes B)(|\Psi>\otimes|\varphi>)&=&
  A|\Psi>\otimes B|\varphi> \\ &=& a|\Psi>\otimes b|\varphi>\\ &=&
  ab|\Psi>\otimes |\varphi>
  \end{eqnarray*}
  And this property applies for the expected value also.}
\end{itemize}

After defining some mathematical tools we can develop the concept
of entanglement on these. Entanglement was first proposed by
Einstein, Podolsky and Rosen as a paradox that implies
incompleteness of Quantum Theory. Entangled states are special
types of states that are defined by only one state function even
if there are more than one state. As an example, one can separate
the ensemble $a|01>+b|00> = |0>\otimes (a|1> + b|0>)$ and
separately apply operators on each state but on an entangled state
such as $\frac{1}{\sqrt 2} |00> + \frac{1}{\sqrt 2} |11>$, there
is no way to separate these states. This immediately results in
the conclusion that if two quantum states are entangled, and if
one of them is measured by applying an operator, say $A$ (or more
explicitly the operator $A \otimes I$), the result of a
measurement by the same operator would be known without
measurement even if they are many light years apart from each
other. This situation, in turn, might seem to mean violating rules
of relativity stating that information can not be propagated
faster than light and there is a paradox according to Einstein,
Podolsky and Rosen. This paradox was resolved by Bell in 1960's
that these states really appear in nature and there is no
violation of relativity since one should send the way of measuring
one state if the other states are supposed to give the same
information.

Later in 1990's EPR states proved themselves to be very valuable,
one can use them to create some type of quantum parallelism
\cite{shor, deutsch} such that algorithms that are proved to be
exponential in time in classical computers were evaluated in
polynomial time in a quantum computer. Moreover entangled states
enable to teleport a state from one place to another by simply
evolving one entangled state in the sender's side so that
receiver's state also evolves giving out the teleported state as a
result.

Now, we will define the entangled states and the operators we will
use in our discussion of networks. We will use the projection
operators $P_0$ and $P_1$ and the entangled state
$|\Psi>=\frac{1}{\sqrt 2}(|00>+|11>)$. If we did the measurement
on the sender's qubit we will have
\begin{equation}
(P_0 \otimes I)|\Psi> = \frac{1} {\sqrt 2} |0>|0>=|\varphi>
\end{equation}
as a result of measurement on one entangled state. An expected
value of $\frac {1} {2}$ will be the output as a result.

And the measurement on the other entangled state after this
evolution gives
\begin{equation}
(I \otimes P_0)|\varphi> = \frac {1} {\sqrt 2} |0>|0>,
\end{equation}
the same result. Note that if $P_1$ is applied on $|\varphi>$ the
result would be 0.

Therefore, we will fix the following convention. Assume that two
nodes A and B in a network have corresponding entangled pairs with
them. If the sender A detects that it is completely isolated from
the network it does a measurement on its qubit by the operator
$P_0$ outputting an expected value of $\frac 1 2$ as a result of
his measurement. When the receiver B sees that no information is
coming via classical channels it does an operation of $P_1$ on its
entangled state. If the node A is isolated node B will get a 0 as
a result, if A is still connected to the network it does no
operation on the system and B will get a result of $\frac 1 2$
from the measurement. This can be also stated in terms of the base
states that in this configuration you either get the two-bit
result 00 from measurement of this two qubit system where the
measurement result is nonzero or you get nothing which means a
zero as output.

\section{Application of entanglement to the counting to infinity
problem}

Entanglement may be applied in many different ways to a classical
network. We will now study these applications, and then compare
them with each other. During these comparisons, we will assume
that the entangled states can be transported via a quantum channel
so that nodes can share them.

One application may be implementing entangled states(there may be
a lot of pairs shared between two nodes) between neighboring nodes
of a selected node A and the distant gateways. When that node A
goes down, neighboring node learns it and then sets its entangled
state accordingly and the distant gateway periodically measures
its entangled state. Since count to infinity procedure has already
begun near the node A, this gateway is aware of the situation and
it may guarantee that the problem does not pass through it to
outer networks and stays in a limited area so that if split
horizon with poisoned reverse is still active the resolution of
the problem becomes faster. This can also help the distance vector
routing to be used in much larger networks since count to infinity
problem is guaranteed to be limited in a smaller area if it occurs
and split horizon with poisoned reverse can handle it up to some
acceptable level. Here the infinity will be represented by the
number of hops to reach the distant gateway that knows the
situation at node A plus 1. This system requires periodical
transportation of entangled pairs since the distant gateway
periodically consumes the states it has by measuring them to learn
if node A has detected any problems around itself. This scenario
is so simple but even in this case the complexity of informing a
distant node about a change in topology becomes in $O(1)$ time.
Variants of this type of informing a change in topology may be
continuously applied among nodes but this also adds an overhead of
qubit exchange on the network. However, this exchange does not
significantly affect the performance since we only send 0 or 1
data and qubits are independent of each other due to the fact that
each of them is discretely entangled and this makes the quantum
channel error tolerant, i.e. if one or two qubits are mistakenly
measured on the way, they become useless but this does not affect
the other qubits and this is not a significant source of
unreliability.

Another application that is more sophisticated is the exchange of
entangled qubits generated at a node while exchanging the distance
vector routing data between neighbors. When one node updates its
table, it also gets corresponding entangled qubits generated by
its neighbors. This procedure can be explained on a simple network
lined up on a straight line which has the morphology
\begin{equation*}
A----B----C-----D----E-- \cdots
\end{equation*}

In this system each node exchanges entangled bits for each entry
of the distance vector routing table. This means that when a node
gets data from a neighbor and decides that neighbor has the
shortest path to some other node, it also gets entangled qubits
generated by that neighboring node. When it needs entry exchange
next time, it also measures the entangled bits it has for
detecting any change in morphology if the neighboring node that
has the other qubit of the entangled pair already detected it.
Therefore, for the above network system each node exchanges
entangled qubits for each of its entries and in each exchange time
these qubits are measured. If node A goes down, it immediately
measures its qubit with an appropriate operator and when node B
requests data for routing table exchange, it gets no answer from
node A and it immediately measures its entangled qubit
corresponding to this entry and learns that node A is not
reachable anymore. Then node B looks for another route for
reaching node A and it detects that node C has a route to node A
without detecting that node C's route passed through it. Node B
still remembers  the result of its measured state for entry A and
when it exchanges entry data between node C it also sends some
entangled qubits  for the corresponding entry of node A to node C.
And since node B knows from its former measurement that node A has
gone down it immediately measures entangled qubits it has for
entry A to state that node A has gone down. Note that at this time
a loop has just begun and count to infinity problem has just
showed itself, i.e. it occurred in the system. After a set of
exchanges between neighboring nodes for the entry of node A, the
node C again looks for its neighbors for a path to node A and
before it gets entry data from node B, it measures the entangled
bits it has for entry A and learns that node B says that node A is
blocked on its side and any path data coming from node A for path
B is not reliable. Therefore node C then learns that node A is not
reachable via node B and looks for other neighbor for alternative
paths to node A. In this way, during each exchange each node first
measures its entangled qubits shared with its neighbor for an
entry in routing table and according to this data it trusts in the
metric its neighbor informs it for another node. Thus the system
quickly and asynchronously(i.e., no periodical measurement of
qubits, they are measured just before each exchange of routing
data and generated and exchanged while exchanging the routing
data) collapses to a stable system with its new network topology.
The protocol defined above can be applied on more complex networks
than the above example and it prevents the count to infinity
problem in an efficient way when it occurs.

\section{Conclusions}

In this paper we briefly defined the distance vector routing
algorithm characteristics and examined possible causes and results
of count to infinity problem. We investigated the advantages,
disadvantages and limitations of the mostly used methods to avoid
count to infinity problem or recover from it. These methods are
split horizon and poisoned reverse which can be used together
also. The reduction of count to infinity problem to halting
problem let us examine the problem from the point of computation
complexities and a decision problem and we proposed to change the
hardware to solve the problem in order to have a more robust
network algorithm without significantly increasing the complexity
of the distance vector routing algorithm. For this reason a
statistical and nondeterministic theory of computation, the
quantum computation theory is used to develop such algorithms. The
novel states called entangled states enable the network node to
communicate with themselves without depending on the network
connectivity in the case of network topology change, which is
possible if a measurement protocol of states is established.
Moreover the time complexity of learning any change(or
equivalently, making a measurement on an entangled state) between
any node with any amount of distance between them is $O(1)$ in the
case of a network topology change, which significantly increases
the size and quality of service of the network on which distance
vector routing can be applied.

\end{document}